\documentstyle[aps,prl,multicol,epsf]{revtex}

\def\R{\hbox{\rm I \kern-5pt R}}

\title{Locality and reality revisited}

\author{Adrian Kent${}^{1,2}$} 
\address{${}^1$ Hewlett-Packard Laboratories, Filton Road,\\
Stoke Gifford, Bristol BS34 8QZ, U.K.\\ 
}
\date{February 2002} 

\begin{document} 
\maketitle
\begin{abstract}

Bell gave the now standard definition of a 
local hidden variable theory and showed that such theories
cannot reproduce the predictions of quantum
mechanics without violating his ``free will'' criterion:
experimenters' measurement choices can be assumed
to be uncorrelated with properties of the measured system prior to
measurement.  

An alternative is considered here: a probabilistic
theory of hidden variables underlying
quantum mechanics could be statistically local, 
in the sense that it supplies global configuration probabilities
which are defined by expressions involving only local terms.   
This allows Bell correlations without relying on {\it either} a
conspiracy theory in which prior common causes correlate  
the system state with experimenters' choices {\it or} 
a reverse causation principle in which experimenters' choices 
affect the earlier system states.  In particular, there is no
violation of the free will criterion.  
It gives a different perspective on Bell correlations,
in which the puzzle is not that apparently non-local
correlations should emerge from rules involving local quantities, 
but rather that we do not see more general non-local correlations
that allow paradox-immune forms of superluminal signalling.   
\end{abstract}
\vskip 5pt
${}^2$ On leave from DAMTP, University of Cambridge, 
Silver Street, Cambridge CB3 9EW, U.K. 


\section{A Lament} 

Many important
concepts in quantum foundations are misleadingly named:

\begin{itemize}

\item The {\it measurement problem} is not the
problem of understanding precisely what a measurement is --- something
which needn't be, and probably isn't, specified in the fundamental
theory of nature --- but the problem of finding a precise mathematical
or physical characterisation of the sample space on which quantum
probabilities are defined.  

\item {\it Hidden variables} are hidden in the
sense that the quantum formalism conceals them, but, as Bell
emphasized, certainly should not be supposed to be hidden from us,
since they are meant to define the reality we experience.  And Bell's own
preferred term, {\it beable}, has a metaphysically confusing
connotation of a not necessarily actualised potential for existence. 
This would seem to make sense for the values of suitable 
variables --- only one value is
realised out of many possibilities --- but not for the 
variables themselves: nonetheless, Bell intended the latter.  
Better terms for quantities parametrising the sample space which 
solves the measurement problem would be {\it existents} 
(Shimony's term\cite{shimonyreply}) 
or perhaps {\it real variables}.   

\item Bell's {\it free will criterion}\cite{bellfreewill}
  conflicts with some widely held
  views of what it means to be able to incorporate free will into a
  physical theory --- and Bell's response to critics explains that he
  did not actually intend to express a view on the problem of free
  will.  Free will, as Bell used it, is no more than a convenient
  shorthand term.  A theory satisfies the free will criterion if it
  allows us to justify treating the measurement settings as
  independent variables to those characterising the state of the
  microscopic system (typically, a pair of separated entangled
  particles) being measured.  This need not mean that the theory
  admits your, or anybody's, favourite notion of free will.
  Conversely, as Price emphasizes\cite{price}, a theory which fails to
  satisfy the free will criterion may nonetheless be compatible with
  sensible definitions of free will.

\item Finally, {\it quantum non-locality} is a confusingly oxymoronic phrase. 
Quantum field 
theories --- the best version of quantum theory we have --- do 
satisfy a precise locality condition: operators with
support in space-like separated regions commute.  
What the term is (or should be) meant to signify
is that quantum theory is not locally causal
under Bell's definitions --- but {\it quantum non-(local causality)} 
is too clumsy.  
\end{itemize} 

What to do?  Using non-standard terms, or inventing new ones, makes
life difficult for readers.  So I will use the standard terms, with
these criticisms understood.  

\section{Some motivation} 

It seems to me that all our difficulties in quantum foundations stem
first from the measurement problem, and then from the fact that the 
obvious ways of tackling the measurement problem are hard to 
reconcile with special relativity and/or reality.  

Probably this is because quantum theory is wrong, and the
insolubility of the measurement problem is a symptom.  Just possibly,
another interpretation of quantum theory might do the trick.  

Either way, there's a case for carefully re-examining the conventional lines
of argument, looking for gaps and trying to find scope for something
new.  Maybe there's also a case for giving up and waiting for 
experiment to give us a clue, but I'll resist that temptation 
here.  Instead, I want to describe a different way of thinking
about hidden variables, which at least sheds a different light on
Bell correlations.    

\section{Lattice models in statistical physics} 

To statistical physicists, lattice models are mathematical 
idealizations.  Many quantities of interest in statistical 
physics are relatively insensitive to the details of the microphysics,
and many distinct systems and models may belong to the
same universality class, with essentially the same properties.
So, if a well chosen lattice model reproduces the bulk 
properties and correlation functions observed in an experimental
sample, this need not imply that the model even approximately 
describes the system's microphysics.  

I want here to use some standard definitions of lattice statistical
physics in an unfamiliar way.  First, I note that
lattice-like models make sense as fundamental theories in their own
right, rather than as phenomenological models attempting to capture
some of the properties of a more fundamental theory.  Then I want to
consider the possibility that something rather like a lattice model
could be a fundamental theory of microphysics, giving a hidden
variable theory explanation of Bell correlations.

Consider the two-dimensional Ising model on a square lattice (Figure
1).  A ``spin'' variable $s_i \in \{ -1, 1 \}$ is attached to
each site $i$.  The Hamiltonian
$$
H = \sum_{ (i, j) ~ {\rm nearest~neighbours}} (1 - s_i s_j ) \, , 
$$ 
includes only interactions between horizontally or vertically
adjacent spins.  For
any possible configuration of spins  
$S = \{ s_i : i~ {\rm a~ lattice~ site} \}$, we have a 
{\it probability weight}
$$
p(S) = \exp ( - C H )   \, , 
$$ 
where $C$ is a constant (at fixed temperature).  
The idealised system represented by the Ising model will always be
found in some legitimate spin configuration, and the probability of finding
it in the configuration $S$ is 
$$
{\rm Prob} (S ) = {{ p (S) } \over { \sum_{{\rm all~ configurations~} S'}
  p(S') }} \, . 
$$

Note that the model is thus {\it statistically local}, in the sense
that if two spin configurations $S$ and $S'$ differ only on a region
$R$, the ratio 
$$
{ {\rm Prob} (S )} \over {{\rm Prob} (S' )} 
$$
can be calculated purely from the values of the $s_i$ and $s'_i$ 
within the region $R$, together with those of their nearest  
neighbours.  That is, if $S$ and $S'$ differ only locally, 
the ratio of their probabilities can be calculated from local
quantities.   

\setlength{\unitlength}{3947sp}%
\begingroup\makeatletter\ifx\SetFigFont\undefined%
\gdef\SetFigFont#1#2#3#4#5{%
  \reset@font\fontsize{#1}{#2pt}%
  \fontfamily{#3}\fontseries{#4}\fontshape{#5}%
  \selectfont}%
\fi\endgroup%
\begin{picture}(6024,5779)(1189,-7328)
\thinlines
\put(2401,-3961){\circle*{150}}
\put(2401,-2761){\circle*{150}}
\put(3601,-3961){\circle*{150}}
\put(3601,-2761){\circle*{150}}
\put(3601,-5161){\circle*{150}}
\put(2401,-5161){\circle*{150}}
\put(2401,-3961){\circle*{150}}
\put(2401,-2761){\circle*{150}}
\put(3601,-3961){\circle*{150}}
\put(3601,-2761){\circle*{150}}
\put(3601,-5161){\circle*{150}}
\put(2401,-5161){\circle*{150}}
\put(2401,-3961){\circle*{150}}
\put(2401,-2761){\circle*{150}}
\put(3601,-3961){\circle*{150}}
\put(3601,-2761){\circle*{150}}
\put(3601,-5161){\circle*{150}}
\put(2401,-5161){\circle*{150}}
\put(2401,-3961){\circle*{150}}
\put(2401,-2761){\circle*{150}}
\put(3601,-3961){\circle*{150}}
\put(3601,-2761){\circle*{150}}
\put(3601,-5161){\circle*{150}}
\put(2401,-5161){\circle*{150}}
\put(4801,-3961){\circle*{150}}
\put(4801,-2761){\circle*{150}}
\put(6001,-3961){\circle*{150}}
\put(6001,-2761){\circle*{150}}
\put(6001,-5161){\circle*{150}}
\put(4801,-5161){\circle*{150}}
\put(6001,-1561){\line( 0,-1){4800}}
\put(4801,-1561){\line( 0,-1){1200}}
\put(4801,-2761){\line( 0,-1){3600}}
\put(1201,-3961){\line( 1, 0){2400}}
\put(3601,-3961){\line( 1, 0){1200}}
\put(4801,-3961){\line( 1, 0){1200}}
\put(6001,-3961){\line( 1, 0){1200}}
\put(3601,-1561){\line( 0,-1){1200}}
\put(3601,-2761){\line( 0,-1){3600}}
\put(2401,-1561){\line( 0,-1){1200}}
\put(2401,-2761){\line( 0,-1){1200}}
\put(2401,-3961){\line( 0,-1){1200}}
\put(2401,-5161){\line( 0,-1){1200}}
\put(1201,-2761){\line( 1, 0){4800}}
\put(6001,-2761){\line( 1, 0){1200}}
\put(1201,-5161){\line( 1, 0){2400}}
\put(3601,-5161){\line( 1, 0){1200}}
\put(4801,-5161){\line( 1, 0){1200}}
\put(6001,-5161){\line( 1, 0){1200}}
\put(2251,-2611){\makebox(0,0)[lb]{\smash{\SetFigFont{12}{14.4}{\rmdefault}{\mddefault}{\updefault}
1
}}}
\put(4651,-2611){\makebox(0,0)[lb]{\smash{\SetFigFont{12}{14.4}{\rmdefault}{\mddefault}{\updefault}
1
}}}
\put(5851,-2611){\makebox(0,0)[lb]{\smash{\SetFigFont{12}{14.4}{\rmdefault}{\mddefault}{\updefault}
1
}}}
\put(3376,-3811){\makebox(0,0)[lb]{\smash{\SetFigFont{12}{14.4}{\rmdefault}{\mddefault}{\updefault}
1
}}}
\put(2251,-5086){\makebox(0,0)[lb]{\smash{\SetFigFont{12}{14.4}{\rmdefault}{\mddefault}{\updefault}
1
}}}
\put(2251,-3811){\makebox(0,0)[lb]{\smash{\SetFigFont{12}{14.4}{\rmdefault}{\mddefault}{\updefault}
1
}}}
\put(3376,-2611){\makebox(0,0)[lb]{\smash{\SetFigFont{12}{14.4}{\rmdefault}{\mddefault}{\updefault}
-1
}}}
\put(4576,-3811){\makebox(0,0)[lb]{\smash{\SetFigFont{12}{14.4}{\rmdefault}{\mddefault}{\updefault}
-1
}}}
\put(5776,-3811){\makebox(0,0)[lb]{\smash{\SetFigFont{12}{14.4}{\rmdefault}{\mddefault}{\updefault}
-1
}}}
\put(3451,-5086){\makebox(0,0)[lb]{\smash{\SetFigFont{12}{14.4}{\rmdefault}{\mddefault}{\updefault}
1
}}}
\put(4576,-5086){\makebox(0,0)[lb]{\smash{\SetFigFont{12}{14.4}{\rmdefault}{\mddefault}{\updefault}
-1
}}}
\put(5776,-5086){\makebox(0,0)[lb]{\smash{\SetFigFont{12}{14.4}{\rmdefault}{\mddefault}{\updefault}
-1
}}}
\put(1726,-7261){\makebox(0,0)[lb]{\smash{\SetFigFont{14}{16.8}{\rmdefault}{\mddefault}{\updefault}
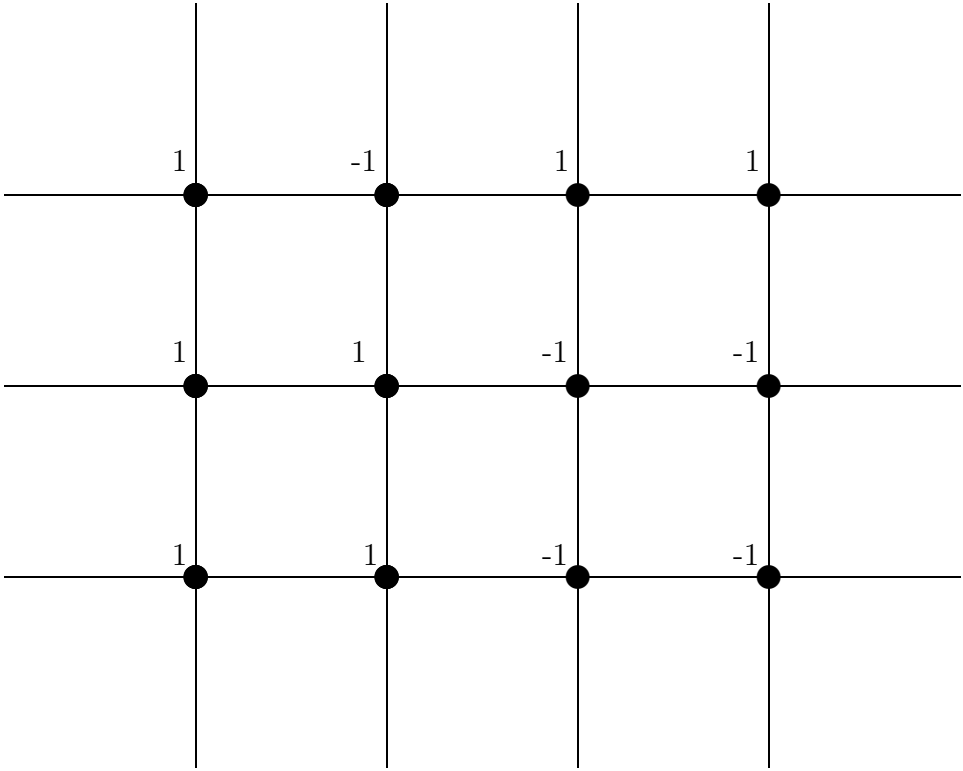
Figure 1: An Ising model spin configuration
}}}
\end{picture}
\vskip10pt
Now, the main point.  {\it The Ising model, as just described,
could be interpreted as a fundamental theory.}  Imagine a world 
in which, say, a particular type of crystal has the property 
that its surfaces are perfect square arrays of atoms, in 
which each atom has an asymmetric feature which define
an orientation, and in which the orientation is always perpendicular
to the surface, either pointing outward ($+1$) or inward ($-1$).  
Imagine further that the orientations are stable --- we cannot
change them --- and can be examined by powerful microscopes, 
and that a plentiful supply of crystals is available.   

Then it makes sense, as a hypothesis, to suppose that 
the atomic orientations are 
randomly configured, with probabilities defined by the Ising
model (on the finite lattice defined by the crystal surface).  
The theory can be tested: examining large numbers of crystals 
will reveal whether or not the data are statistically consistent with
the theory.   And it need not be a phenomenological theory; it could
be fundamental.  It could simply be that nature has fixed these 
unchanging orientation configurations, not through some underlying
dynamics whose end result is described by the Ising model, but 
from the beginning, by a random choice which (like the randomness
of quantum measurements, according to the orthodox view) has no underlying
explanation.  

Obviously, this imaginary Ising-world is not much like ours, in which
crystals form and decompose, atoms don't behave classically, their
configurations and spin states aren't fixed, and everything is (as far
as we presently understand) ultimately described by quantum dynamics.
But conceptually the Ising theory of crystals in Ising-world makes
sense.  It is testable in much the same way as quantum theory is, and
it includes a simplifying principle --- statistical locality --- which
makes it much easier to test: it suffices to look at orientation
configurations in local regions of crystal surfaces.  And, given our
other assumptions about Ising-world, there is no compelling reason ---
nothing analogous to the measurement problem --- to think Ising theory
must be incomplete.

\section{Minkowski space trajectory models} 

Let us suppose that underlying quantum theory there is some 
theory in which real particles follow definite trajectories 
through spacetime.  To simplify things, take 
space-time to be two-dimensional, with a discrete lattice
structure at microscopic scales (Figure 2). 
The lattice edges are lightlike, 
and so the discussion is restricted to massless particles, which 
travel along the lattice edges.   
Also, for the purposes of this discussion, macroscopic experimental
apparatus will be treated as qualitatively distinct from 
microscopic particles, so that we are allowed to consider 
apparatus set up somewhere in spacetime without having to 
describe it as a collection of microscopic particles, each
following their own trajectories through the lattice.

All of these simplifications are meant to be provisional rather
than fundamental.  
The idea is to set up the standard discussion
of Bell correlations, and then find a new way of looking at them,
without obscuring things with too many technicalities.
\vskip10pt
\setlength{\unitlength}{3947sp}%
\begingroup\makeatletter\ifx\SetFigFont\undefined%
\gdef\SetFigFont#1#2#3#4#5{%
  \reset@font\fontsize{#1}{#2pt}%
  \fontfamily{#3}\fontseries{#4}\fontshape{#5}%
  \selectfont}%
\fi\endgroup%
\begin{picture}(6087,6424)(301,-7403)
\thinlines
\put(976,-6286){\vector(-1, 1){450}}
\put(976,-6286){\vector( 1, 1){450}}
\put(2080,-5234){\line( 1, 1){3394}}
\put(2928,-6082){\line( 1, 1){3394}}
\put(1231,-4385){\line( 1, 1){3394}}
\put(1231,-2688){\line( 1,-1){3394}}
\put(2082,-1842){\line( 1,-1){3394}}
\put(2982,-1017){\line( 1,-1){3394}}
\put(1351,-6211){\makebox(0,0)[lb]{\smash{\SetFigFont{12}{14.4}{\rmdefault}{\mddefault}{\updefault}
x+ct
}}}
\put(301,-6136){\makebox(0,0)[lb]{\smash{\SetFigFont{12}{14.4}{\rmdefault}{\mddefault}{\updefault}
x-ct
}}}
\put(3777,-3537){\circle*{150}}
\put(2928,-4385){\circle*{150}}
\put(2928,-2688){\circle*{150}}
\put(2080,-3537){\circle*{150}}
\put(3777,-1840){\circle*{150}}
\put(4625,-2688){\circle*{150}}
\put(4623,-4399){\circle*{150}}
\put(3758,-5254){\circle*{150}}
\put(5485,-3542){\circle*{150}}
\put(1051,-7336){\makebox(0,0)[lb]{\smash{\SetFigFont{14}{16.8}{\rmdefault}{\mddefault}{\updefault}
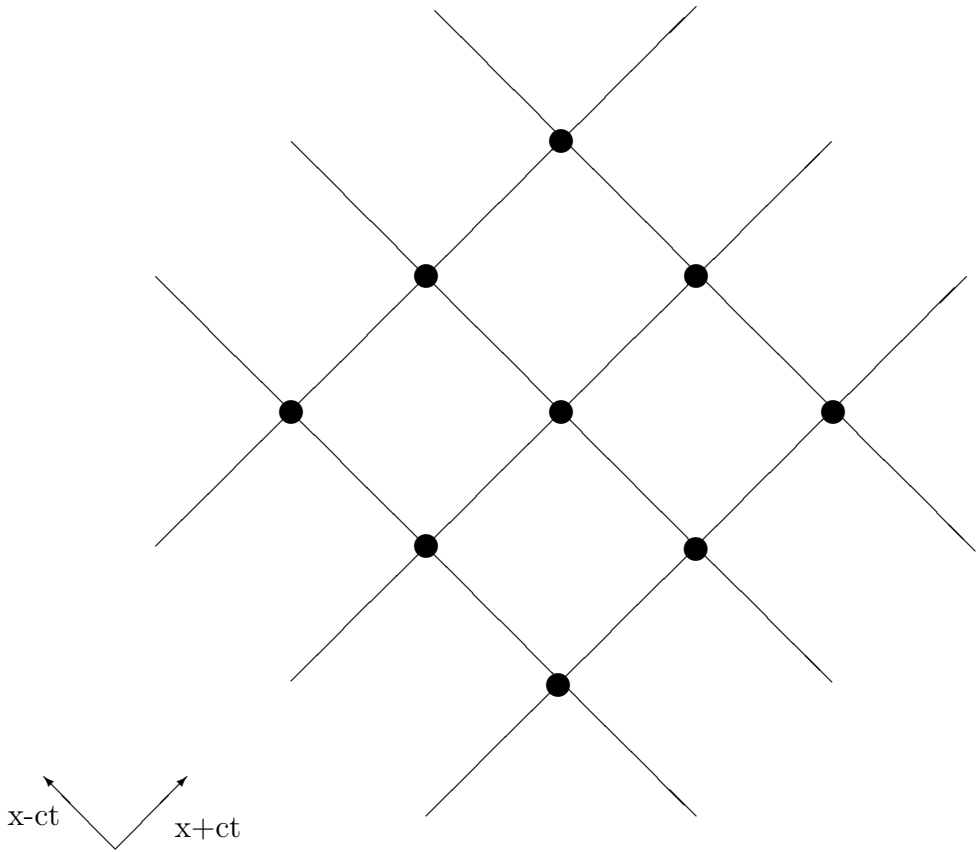
Figure 2:  Two-dimensional Minkowski space trajectory lattice
}}}
\end{picture}
\vskip10pt

Figure 3 shows a standard Bell experiment on a pair of entangled
massless particles, ``photons'', 
with one extra feature --- lightlike trajectories 
along the lattice edges have been included.   
Though this looks like a standard beam path picture, the trajectories
here are hidden variables: the picture is meant to be interpreted
as saying that the photons {\it really} did travel along the paths 
depicted --- something we will eventually have to show is 
consistent with the model we produce.  
The space-time trajectories of the classical measuring apparatuses 
won't affect the discussion, and are omitted.
\vskip10pt
\setlength{\unitlength}{3947sp}%
\begingroup\makeatletter\ifx\SetFigFont\undefined%
\gdef\SetFigFont#1#2#3#4#5{%
  \reset@font\fontsize{#1}{#2pt}%
  \fontfamily{#3}\fontseries{#4}\fontshape{#5}%
  \selectfont}%
\fi\endgroup%
\begin{picture}(6324,4438)(1189,-6287)
\thinlines
\put(4801,-4561){\vector( 1, 1){2700}}
\put(4801,-4561){\vector(-1, 1){2700}}
\put(6451,-2911){\circle{424}}
\put(3151,-2911){\circle{424}}
\put(1801,-5161){\vector( 1, 1){600}}
\put(1801,-5161){\vector(-1, 1){600}}
\put(3901,-4936){\makebox(0,0)[lb]{\smash{\SetFigFont{12}{14.4}{\rmdefault}{\mddefault}{\updefault}
source of entangled photons
}}}
\put(1576,-2986){\makebox(0,0)[lb]{\smash{\SetFigFont{12}{14.4}{\rmdefault}{\mddefault}{\updefault}
~~~~spin~~~~
}}}
\put(1576,-3211){\makebox(0,0)[lb]{\smash{\SetFigFont{12}{14.4}{\rmdefault}{\mddefault}{\updefault}
measurement
}}}
\put(6901,-2911){\makebox(0,0)[lb]{\smash{\SetFigFont{12}{14.4}{\rmdefault}{\mddefault}{\updefault}
~~~~spin~~~~
}}}
\put(6901,-3136){\makebox(0,0)[lb]{\smash{\SetFigFont{12}{14.4}{\rmdefault}{\mddefault}{\updefault}
measurement
}}}
\put(3301,-6211){\makebox(0,0)[lb]{\smash{\SetFigFont{14}{16.8}{\rmdefault}{\mddefault}{\updefault}
Figure 3: A Bell experiment
}}}
\put(2251,-5086){\makebox(0,0)[lb]{\smash{\SetFigFont{12}{14.4}{\rmdefault}{\mddefault}{\updefault}
x+ct
}}}
\put(1201,-5086){\makebox(0,0)[lb]{\smash{\SetFigFont{12}{14.4}{\rmdefault}{\mddefault}{\updefault}
x-ct
}}}
\end{picture}
\vskip10pt
To further simplify things --- since we cannot use photon polarizations
in two dimensions --- let us assume
that the measurements are perfect von Neumann measurements of
an internal degree of freedom, ``spin'', with the  
photons initially in the singlet state of a pair of spin $1/2$ 
particles.\footnote{The scare quotes around ``photon'' and ``spin'' are
left implicit from now on.}   According to quantum theory, these von Neumann
measurements do not affect
the form of the photon wave packet, assuming (let's) that the spatial 
and spin degrees of freedom are unentangled.
It seems reasonable to postulate that, in the hidden variable model we are 
building, the photon trajectories will also be unaffected by such 
measurements.   For instance, after a measurement of spin in the 
measurement basis defined by vector ${\bf a}$, the apparatus records the
answer ${\bf a}$ or ${\bf -a}$, 
the photon emerges in the corresponding spin eigenstate, and 
its trajectory continues unaltered.  

The aim now is to reproduce Bell correlations by a 
statistically local lattice-like model involving photon trajectories, 
while respecting Bell's free will criterion.  We will want to
attach further hidden variable degrees of freedom to the 
photon trajectories, so that the spin measurements become
relevant to the hidden variable model.  
The free will criterion requires that 
no hidden variables at points before the measurements should 
be correlated with the measurement choices.  This, together with
locality, means that the photons' states before the measurements 
cannot explain the correlations: we need rules 
involving events {\it after} the measurements.   

We can ensure a relevant post-measurement event if the photons
are forced to meet again, and {\it that}
can be done, without violating statistical locality, as follows.
First, we give a small but non-zero probability weight to 
trajectory segments in which a photon's direction switches: 
this allows the two photons to meet after the measurements. 
Second, we choose rules that assign probability zero to the event 
that the photons propagate forever, and non-zero probability
weight to the event that two photon trajectories annihilate
when they meet.  Together these ensure that photons {\it must}
eventually undergo pair annihilation. 
The first of these can be arranged by assigning probability weight
of $(1- \epsilon )$  to a vertex in which the photon continues in the
same direction, and weight $\epsilon$ to a vertex
in which it switches (Figure 4); here $\epsilon > 0$ is chosen to 
be small enough that $ ( 1 - \epsilon )^N \approx 1$, where $N$
is the number of lattice links a photon would traverse in the
largest scale Bell experiment carried out to date.    

\vskip10pt
\setlength{\unitlength}{3947sp}%
\begingroup\makeatletter\ifx\SetFigFont\undefined%
\gdef\SetFigFont#1#2#3#4#5{%
  \reset@font\fontsize{#1}{#2pt}%
  \fontfamily{#3}\fontseries{#4}\fontshape{#5}%
  \selectfont}%
\fi\endgroup%
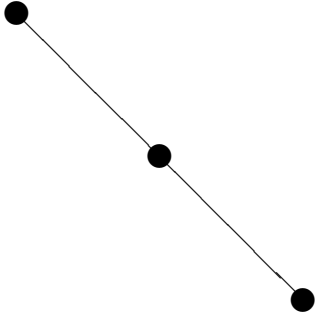
\begin{picture}(6158,3474)(976,-6227)
\thinlines
\put(7051,-2836){\line(-1,-1){900}}
\put(6151,-3736){\line( 1,-1){900}}
\put(2071,-6151){\makebox(0,0)[lb]{\smash{\SetFigFont{14}{16.8}{\rmdefault}{\mddefault}{\updefault}
Figure 4: Probability Weights for Free Progagation
}}}
\put(5476,-5326){\makebox(0,0)[lb]{\smash{\SetFigFont{14}{16.8}{\rmdefault}{\mddefault}{\updefault}
Probability weight =  $\epsilon$
}}}
\put(976,-5326){\makebox(0,0)[lb]{\smash{\SetFigFont{14}{16.8}{\rmdefault}{\mddefault}{\updefault}
Probability weight =  1 - $\epsilon$ 
}}}
\put(1651,-2836){\line( 1,-1){1800}}
\put(7051,-4636){\circle*{150}}
\put(1651,-2836){\circle*{150}}
\put(3451,-4636){\circle*{150}}
\put(2551,-3736){\circle*{150}}
\put(7051,-2836){\circle*{150}}
\put(6151,-3736){\circle*{150}}
\end{picture}
\vskip10pt
In choosing rules for a pair annihilation, we need to make sure that 
the originally entangled photons meet and mutually annihilate
each other, and not other photons.  This we do by introducing a hidden
variable label which only the entangled pair share, and allowing
annihilation only between pairs which share the same label.    
Finally, we enforce the correct correlations by fiat, by adding
a further label which records a photon's measurement outcome, and
by making the probability weight for the annihilation event
depend appropriately on these labels (Figure 5).  
\vskip10pt
\setlength{\unitlength}{3947sp}%
\begingroup\makeatletter\ifx\SetFigFont\undefined%
\gdef\SetFigFont#1#2#3#4#5{%
  \reset@font\fontsize{#1}{#2pt}%
  \fontfamily{#3}\fontseries{#4}\fontshape{#5}%
  \selectfont}%
\fi\endgroup%
\begin{picture}(5108,3803)(1576,-6481)
\thinlines
\put(4801,-2761){\line( 1,-1){1800}}
\put(4801,-2761){\line(-1,-1){1800}}
\put(2026,-6481){\makebox(0,0)[lb]{\smash{\SetFigFont{14}{16.8}{\rmdefault}{\mddefault}{\updefault}
\qquad \qquad     Bell Particles after Measurements
}}}
\put(2071,-6151){\makebox(0,0)[lb]{\smash{\SetFigFont{14}{16.8}{\rmdefault}{\mddefault}{\updefault}
Figure 5: Probability Weight for Recombining 
}}}
\put(1801,-3361){\makebox(0,0)[lb]{\smash{\SetFigFont{12}{14.4}{\rmdefault}{\mddefault}{\updefault}
particle from pair i
}}}
\put(1801,-3361){\makebox(0,0)[lb]{\smash{\SetFigFont{12}{14.4}{\rmdefault}{\mddefault}{\updefault}
particle from pair i after 
}}}
\put(1576,-3586){\makebox(0,0)[lb]{\smash{\SetFigFont{12}{14.4}{\rmdefault}{\mddefault}{\updefault}
measurement result a
}}}
\put(6201,-3361){\makebox(0,0)[lb]{\smash{\SetFigFont{12}{14.4}{\rmdefault}{\mddefault}{\updefault}
particle from pair j after  
}}}
\put(6201,-3586){\makebox(0,0)[lb]{\smash{\SetFigFont{12}{14.4}{\rmdefault}{\mddefault}{\updefault}
measurement result b 
}}}
\put(3001,-5461){\makebox(0,0)[lb]{\smash{\SetFigFont{14}{16.8}{\rmdefault}{\mddefault}{\updefault}
Probability 
weight = $\delta_{ij} \, ( 1 -  a.b) /2 $ 
}}}
\put(3001,-4561){\circle*{150}}
\put(4801,-2761){\circle*{150}}
\put(6601,-4561){\circle*{150}}
\end{picture}
\vskip10pt
To summarise: we've arranged rules so that the entangled photons from
pair $i$ do
the following: (i) with high probability, propagate linearly from
source to detectors, (ii) after measurement, each eventually undergo 
one or more changes of direction, (iii) with probability one, 
eventually meet up and annihilate.  (Figure 6) 
\vskip10pt
\setlength{\unitlength}{3947sp}%
\begingroup\makeatletter\ifx\SetFigFont\undefined%
\gdef\SetFigFont#1#2#3#4#5{%
  \reset@font\fontsize{#1}{#2pt}%
  \fontfamily{#3}\fontseries{#4}\fontshape{#5}%
  \selectfont}%
\fi\endgroup%
\begin{picture}(6012,9511)(2401,-8903)
\thinlines
\put(8401,-5461){\line(-1, 1){1800}}
\put(6601,-3661){\line( 1, 1){900}}
\put(7501,-2761){\line(-1, 1){1800}}
\thicklines
\put(6001,-7861){\line(-1, 1){525}}
\thinlines
\put(6676,-7186){\line( 1, 1){1725}}
\thicklines
\put(6001,-7861){\line( 1, 1){525}}
\thinlines
\put(8401,-5461){\line(-1, 1){1800}}
\put(6601,-3661){\line( 1, 1){900}}
\put(7501,-2761){\vector(-1, 1){1800}}
\put(5401,-7261){\line(-1, 1){2700}}
\put(2701,-4561){\line( 1, 1){600}}
\put(3301,-3961){\line(-1, 1){300}}
\put(3001,-3661){\vector( 1, 1){2700}}
\put(7051,-7486){\makebox(0,0)[lb]{\smash{\SetFigFont{12}{14.4}{\rmdefault}{\mddefault}{\updefault}
result b
}}}
\put(2401,-8836){\makebox(0,0)[lb]{\smash{\SetFigFont{14}{16.8}{\rmdefault}{\mddefault}{\updefault}
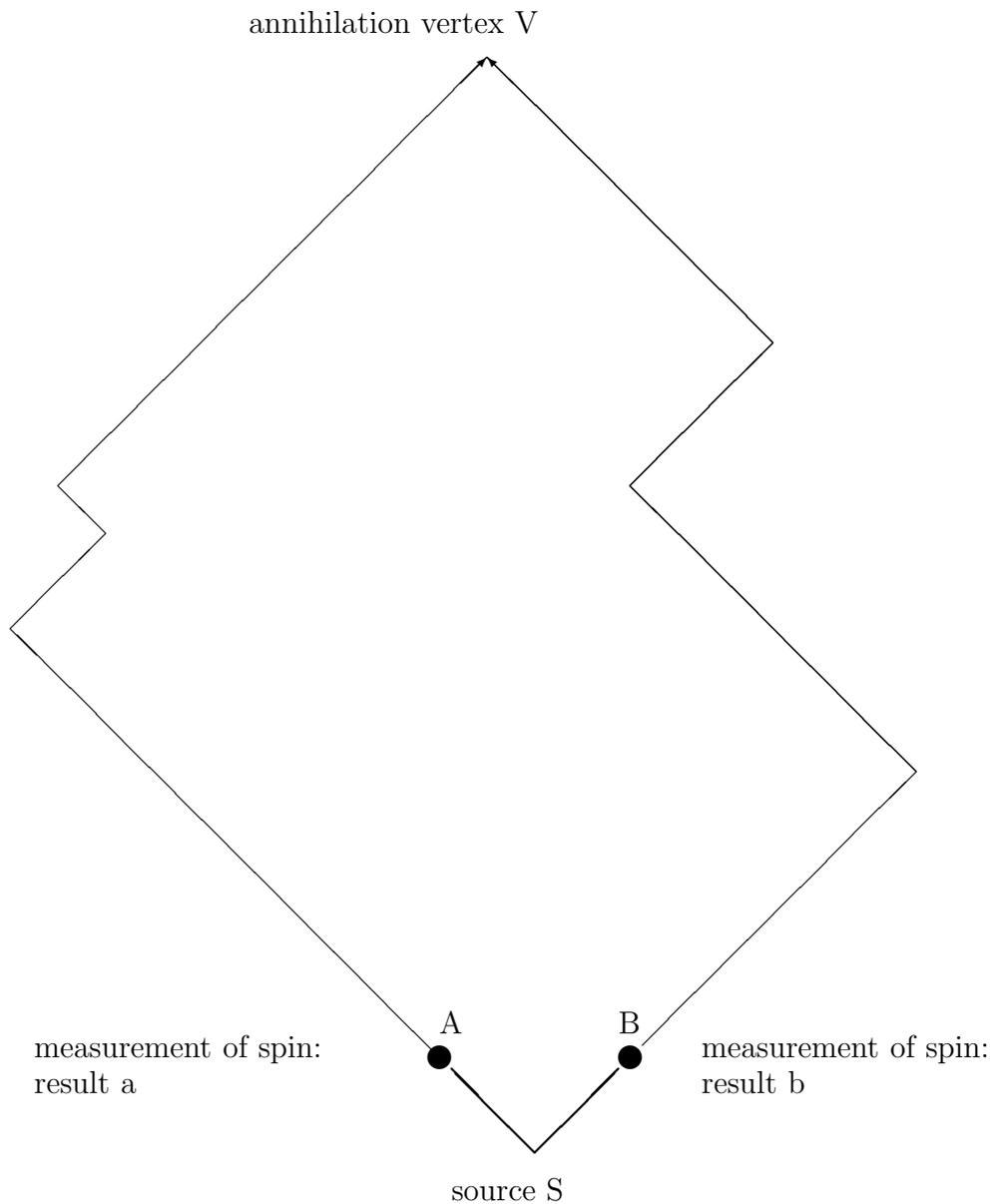
Figure 6: a typical trajectory description of a Bell experiment in the model
}}}
\put(5401,-7111){\makebox(0,0)[lb]{\smash{\SetFigFont{12}{14.4}{\rmdefault}{\mddefault}{\updefault}
A
}}}
\put(6526,-7111){\makebox(0,0)[lb]{\smash{\SetFigFont{12}{14.4}{\rmdefault}{\mddefault}{\updefault}
B
}}}
\put(4201,-811){\makebox(0,0)[lb]{\smash{\SetFigFont{12}{14.4}{\rmdefault}{\mddefault}{\updefault}
annihilation vertex V
}}}
\put(5476,-8161){\makebox(0,0)[lb]{\smash{\SetFigFont{12}{14.4}{\rmdefault}{\mddefault}{\updefault}
source S
}}}
\put(2526,464){\makebox(0,0)[lb]{\smash{\SetFigFont{12}{14.4}{\rmdefault}{\mddefault}{\updefault}
The trajectory  from A to V carries hidden variables i (pair label) and a (measurement result)
}}}
\put(2526,164){\makebox(0,0)[lb]{\smash{\SetFigFont{12}{14.4}{\rmdefault}{\mddefault}{\updefault}
The trajectory from B to V carries hidden variables i (pair label) and b (measurement result)
}}}
\put(2526,-136){\makebox(0,0)[lb]{\smash{\SetFigFont{12}{14.4}{\rmdefault}{\mddefault}{\updefault}
The trajectories from S to A and B carry hidden variable i (pair label)
}}}
\put(2851,-7486){\makebox(0,0)[lb]{\smash{\SetFigFont{12}{14.4}{\rmdefault}{\mddefault}{\updefault}
result a 
}}}
\put(2851,-7261){\makebox(0,0)[lb]{\smash{\SetFigFont{12}{14.4}{\rmdefault}{\mddefault}{\updefault}
measurement of spin:
}}}
\put(7051,-7261){\makebox(0,0)[lb]{\smash{\SetFigFont{12}{14.4}{\rmdefault}{\mddefault}{\updefault}
measurement of spin:
}}}
\put(6601,-7261){\circle*{150}}
\put(5401,-7261){\circle*{150}}
\end{picture}
\vskip10pt

\section{Discussion}

All we've done, of course, is rewrite the standard quantum
correlations into ad hoc rules, in such a way that we can 
formally attach hidden variables characterising the measurement
results to the photon paths after measurement, and formally 
account for the correlations as deriving from a rule attached 
to a pair annihilation event that must take place in the far
future.  What was the point of that?  

The point, as I see it, is to underline how difficult it is 
to draw absolutely watertight conclusions about the form
of realist theories underlying quantum theory, even if we 
accept that such theories should respect 
special relativity and should incorporate some form of
locality, and even after decades of analysis of the implications
of Bell experiments.  To recapitulate:  

We can agree with Bell that any 
theory which postulates 
common causes in the past which systematically affect experimenter
and system, in such a way as to always produce Bell correlations, 
is likely to be unbelievably conspiratorial.   But, as 
Price has so carefully and persuasively argued\cite{price}, this leaves
open the possibility of theories which invoke some form of advanced action 
principle, by which experimenters' choices affect the earlier state of
the system.   As Price notes, although such theories would violate
Bell's free will
criterion, they would not necessarily rule out free will {\it per se}, at least
according to some reasonable definitions.  And they would not contradict
the notion of effective freedom which underlay Bell's definition, nor
invoke the same type of implausible conspiracy as a common
cause explanation.  

Still, there is (at the very least) an aesthetic case for the free 
will criterion.  But even if the criterion is accepted, the discussion
isn't closed.  The models described above respect the free
will criterion and they invoke no complex common cause hypothesis. 
They incorporate a respectable form of locality. 
Yet they give --- in the simple case considered --- a consistent
explanation of Bell experiments.   

That said, there is, of course, much to be said against the sort
of ideas outlined above.  The rules proposed are entirely ad hoc, 
transparently constructed precisely in order to reproduce quantum
correlations.  They could be generalised to cover more complicated
entanglement experiments, but only --- it would seem --- at the price
of introducing even more ad hoc and unwieldy rules.  They require,
in particular, that every hidden particle trajectory carries a record of
its past history --- which particles it was originally entangled
with, and which measurements were carried out on it --- arbitrarily 
far into the future.  It is hard to see, methodologically speaking,
what the point of pursuing this idea would be, unless some independent
principle could be found to explain why the annihilation vertex  
weights should take the form necessary to reproduce quantum
correlations.   And where could such a principle come from?

In fact, any type of correlation, including those which allow superluminal
signalling, could consistently be attached to the annihilation
vertex.  Why should such correlations be excluded?  If we were to take 
the basic idea of these models seriously, we would be led to ask why
the non-local correlations we see in nature are so tame as to 
preclude superluminal signalling.  That this would lead to paradoxes
is not a valid answer.  Consistent probabilistic rules for global 
configurations cannot produce a logical contradiction.  Attempts
to set up a paradoxical causal loop fail.  This is essentially because the
probability of a successful superluminal signal is --- even in
isolation --- always at least slightly less than one, and the 
probability of a sequence of signals succeeding is {\it not}
generally the product of the values of the individual probabilities
of success if the relevant signalling devices were 
isolated\cite{aksignal}.  

Moreover, the model involved several breathtaking simplifications,
which would need to be removed if the idea were to be taken seriously.
(Can we eliminate the need for an underlying lattice, or ensure that
the model has a Lorentz invariant continuum limit?  What about rules
for massive particles, and for other types of measurement?  
And how do we build a coherent model that describes macroscopic
measuring devices in terms of microscopic particles?)   

Nonetheless, despite all these flaws, I think there is some 
interest in this rewriting of the standard story.   It seems to 
show that the conventional wisdom about Bell's ``free will'' 
criterion is not quite right: realist, local and Lorentz invariant theories
can respect the criterion and yet reproduce quantum correlations.  
And it does so by identifying a class of realist models --- 
probabilistic models in which the probability of a configuration 
depends on a product of local terms --- which 
standard discussions of Bell's theorem neglect.  
There are, as just noted, many arguments against taking seriously
the idea that nature is actually described by these models.  But the 
arguments are different from the arguments against 
other unorthodox ideas about hidden variables and Bell correlations, 
and worth thinking through, just for that reason.

\section{Acknowledgements} 

I am very grateful to Tomasz Placek and Jeremy Butterfield for
giving me the opportunity to present this work at a most stimulating
and enjoyable meeting, and for patient editorial
encouragement far beyond the call of duty.  The work was supported in
part by a Royal Society University Research Fellowship, PPARC, and the
European collaboration EQUIP.

\end{document}